\documentclass[10pt,twocolumn]{IEEEtran}
\usepackage{multicol}

\usepackage{color}
\usepackage{soul}
\usepackage{amsmath}
\usepackage{graphics}
\usepackage{setspace}
\usepackage{cite}
\usepackage{latexsym}
\usepackage{float}
\usepackage{epsfig}
\usepackage{multirow}
\usepackage[table,xcdraw]{xcolor}
\usepackage{cite,cases,url}
\usepackage{amssymb}
\usepackage{graphicx}
\usepackage{epstopdf}
\usepackage{balance}
\usepackage{subcaption}
\usepackage{enumerate}
\usepackage{array}
\usepackage{algorithmic}
\usepackage{algorithm}
\usepackage{enumitem}
\usepackage[normalem]{ulem}
\usepackage{longtable}
\usepackage{ragged2e}
\usepackage{booktabs}
\usepackage{tabularx}

\useunder{\uline}{\ul}{}

\newcolumntype{L}{>{\centering\arraybackslash}m{5cm}}
\newcolumntype{K}{>{\centering\arraybackslash}m{6cm}}
\newcolumntype{P}{>{\centering\arraybackslash}m{2.3cm}}
\newcolumntype{M}{>{\raggedright\arraybackslash}m{2cm}}
\newcolumntype{N}{>{\raggedright\arraybackslash}m{2.5cm}}

\begin{document}

\title{
Advancing Experimental Platforms for UAV Communications: Insights from AERPAW'S Digital Twin 
}
\author{
\IEEEauthorblockN{Joshua Moore, Aly S. Abdalla, Charles Ueltschey, An\i l G\"urses , Özgür Özdemir, Mihail L. Sichitiu, Ismail Güven\c{c}, and Vuk Marojevic \\
} 
\normalsize\IEEEauthorblockA{Department of Electrical and Computer Engineering,  Mississippi State University, USA \\
Department of Electrical and Computer Engineering,  North Carolina State University, USA} \\
\normalsize\IEEEauthorblockA{\{jjm702,asa298,cmu32,vm602\}@msstate.edu}, 
\{agurses,oozdemi,mlsichit,iguvenc\}@ncsu.edu
\vspace{-0.8cm}}
\maketitle

\begin{abstract}
The rapid evolution of 5G and beyond has advanced space-air-terrestrial networks, with unmanned aerial vehicles (UAVs) offering enhanced coverage, flexible configurations, and cost efficiency. However, deploying UAV-based systems presents challenges including varying propagation conditions and hardware limitations. While simulators and theoretical models have been developed, real-world experimentation is critically important to validate the research. Digital twins, virtual replicas of physical systems, enable emulation that bridge theory and practice. This paper presents our experimental results from AERPAW’s digital twin, showcasing its ability to simulate UAV communication scenarios and providing insights into system performance and reliability. 
\end{abstract}
\IEEEpeerreviewmaketitle

\section{Introduction}
\label{sec:intro}
The rapid evolution of wireless communication technologies, particularly with the advent of 5G and beyond, has ushered in a new era of interconnected space-air-terrestrial networks. Among these advancements, airborne networks, featuring unmanned aerial vehicles (UAVs), have garnered significant attention due to their potential to extend wireless coverage, provide flexible network configurations, and reduce capital expenditures compared to traditional terrestrial networks. UAVs promise to revolutionize various sectors, including disaster response, agriculture, and delivery services.

However, the deployment and optimization of UAV-based communication systems present several challenges. These include managing varying propagation conditions and 3D coverage, addressing spectrum access and interference, and ensuring reliable performance. To tackle these issues, both theoretical analyses and computer simulations have traditionally been employed. 
There is a growing consensus that field experimentation alongside simulation is indispensable for gaining a comprehensive understanding of UAV communications.

Digital twins (DTs) have emerged as a transformative technology across various fields, providing virtual representations of physical systems that enable accurate simulation, detailed analysis, and optimization. There have been various DT platforms that have been established for advancing wireless communications beyond the current state-of-art research efforts~\cite{villa2024colosseum, polese2024colosseum, testolina2024boston, eldeeb2024digital}. The work presented in~\cite{villa2024colosseum} discusses the potential of the Colosseum testbed as a large-scale wireless network emulator for scalable experiments, artificial intelligence (AI) integration, and comprehensive evaluation of complex wireless scenarios such as vehicular networks, Internet of Things, and 5G networks. Building upon this, in~\cite{polese2024colosseum} the authors have showcased how the Colosseum can be further emphasized for open radio access network (O-RAN) DT. Colosseum O-RAN can be utilized to emulate O-RAN components and interfaces with AI-driven techniques and integrated real-world data to test and optimize RAN configurations and protocols in a virtual environment before real-world deployment. In~\cite{testolina2024boston}, the Boston Twin is introduced as a DT designed for precise ray-tracing simulations in the context of 6G networks. Boston Twin provides highly accurate models of urban environments with ray-tracing algorithms to emulate the propagation of wireless signals that enables further research and development (R\&D) opportunities to optimize 6G network deployments. The authors of~\cite{testolina2024boston} presents the fundamentals and potentials of the DT framework for optical wireless communication to support real-time monitoring, simulation, and optimization with predictive analysis, allowing the system to anticipate and mitigate potential system issues. 
 
The Aerial Experimentation and Research Platform for Advanced Wireless (AERPAW) stands out as a pivotal platform for UAV experimentation~\cite{9625086}. AERPAW's DT, a virtual replica of the physical testbed, provides a robust environment for defining and executing aerial communications experiments, thereby facilitating the validation and refinement of UAV communication technologies before real-world deployment~\cite{aerpawarch}.

In this paper, we show results using AERPAW's vehicle emulator with a custom experiment developed in AERPAW's development environment, or DT. This experiment consists of an aerial base station serving two fixed nodes leveraging AERPAWs vehicle emulation and emulated radio frequency (RF) link. The results highlight the capabilities of the AERPAW testbed in emulating real-world scenarios and provide valuable insights into the performance and reliability of UAV-based advanced wireless communication systems. The use of the vehicle emulator enables detailed analysis of network behavior under controlled conditions, allowing for a deeper understanding of potential challenges and opportunities of cellular communications systems for and with UAVs.

The rest of this paper is organized as follows. Section II details the design and use cases for UAV experimentation in a DT. Section III provides an overview of AERPAW's DT. Section IV describes the experimental procedures and results.  Section V identifies critical research directions and derives the conclusions.

\section{
Digital Twins for UAV Experimentation}
\label{sec:architecture}
DTs have become indispensable in wireless experimentation by offering sophisticated virtual models that replicate the behavior and interactions of 
mobile users in advanced wireless networks. These digital representations facilitate comprehensive emulation and optimizations of wireless communication scenarios, enabling researchers to fine-tune and enhance the performance of wireless systems in a controlled, fully-digital experimental environment. By bridging the gap between theoretical research and practical application, DTs empower researchers to explore and validate innovative solutions, driving advancements in wireless technology and mobile communications.

In the following subsections, we delve into the key components, tradeoffs of DTs with other methods supporting experimentation with various use cases, emphasizing their importance in advancing UAV communications.

\subsection{Key Components for Wireless DTs}

Understanding the essential components of DT platforms is critical for their effective deployment in wireless research. These components, such as RF propagation models and real-time RF data integration, are foundational for creating accurate and reliable experiments. By exploring these elements, we can better appreciate how DTs can emulate complex wireless environments and enhance system performance. Recent literature highlights several essential components required for effective DT platforms in wireless research \cite{alkhateeb2023realtimedigitaltwinsvision}:

\begin{itemize}
    \item \textbf{RF Propagation Models:} simulate the propagation of radio signals through various environments and conditions.
    \item \textbf{Channel Emulation:} replicates the characteristics of wireless channels to assess received signal quality and network performance.
    \item \textbf{Antenna Models:} represent the behavior and performance of antennas in the network.
    \item \textbf{Network Configuration Models:} depict the layout and parameters of network components.
    \item \textbf{Real-Time RF Data Integration:} incorporates live data to ensure that simulations reflect current conditions.
    \item \textbf{RF Performance Analytics:} analyzes performance metrics to evaluate and optimize wireless systems.
    \item \textbf{Scenario and Environment Modeling:} develops and manages virtual scenarios for testing various conditions and configurations.
\end{itemize}

\subsection{Balancing DTs with Other Methods}

While DTs offer significant advantages, it is important to balance them with other research methods, including simulations, experiments in sandbox environments, and experiments in physical testbeds. Each research platform has its strengths and limitations, and combining them ensures comprehensive and reliable results. This section highlights the importance of integrating DTs with other testing methodologies to optimize their effectiveness and applicability.

\begin{itemize}
    \item \textbf{Simulations:} efficiently analyze systems but may not directly apply to real-world scenarios due to implementation constraints. DTs often use Software-in-the-Loop emulators, which operate with the actual software of a production-like system to virtually emulate its functions. However, accurate wireless emulation remains challenging and requires adjustments with real-world data.
    \item \textbf{Sandbox Environments:} offer valuable insights using actual hardware but lack the scalability of digital models. They are useful for testing specific components or configurations but are limited in their scope.
    \item \textbf{Physical Testbeds:} provide precise validation of systems and scenarios but are expensive and time-consuming. They are essential for real-world testing and verification but may not always be practical, especially in early-stage research that pushes the boundaries of existing technology.
\end{itemize}

Considering these tradeoffs is important for leveraging the strengths of DTs while addressing their limitations and ensuring comprehensive and reliable results and conclusions~\cite{gürses2024digitaltwinssupportingai}.

\subsection{Use Cases}
DTs play a versatile role in various aspects of UAV experimentation, from simulation and testing to control and navigation, safety and security, communication, and edge computing. By examining specific use cases, we can see how DTs bridge the gap between theoretical models and practical implementations, contributing to significant advancements in UAV research and development. Some of the DT use cases for UAV-assisted wireless communications include trajectory planning and optimization, cooperative swarm UAV deployments, coexistence aerial and ground communications, and UAV-reconfigurable intelligent surfaces assisted vehicular communications \cite{cheng2024enhancedreinforcementlearningbasedresource}. 
DTs can be used to provide a comprehensive digital replica of complex physical world scenarios to test and optimize these use cases. 
To enable wider coverage area and strong line of sight (LoS) communications with optimized trajectories, physical layer configurations, and resource allocations for aerial nodes providing wireless communications services before the real world deployments.



\section{
\textcolor{black}{AERPAW's Digital Twin Overview}}
\label{sec:testing}
AERPAW's testbed provides a structured lifecycle for experimental research, involving several stages: Development (design and container configuration), Emulation (software model validation in the DT), and Testbed Execution (deployment and execution on the AERPAW testbed)~\cite{aerpawarchhighlevel}. Equipped with a variety of software-defined radios (SDRs) such as Universal Software Radio Peripherals (USRPs), the testbed supports custom waveform and protocol development, facilitating studies on aerial coverage, link quality, interference, and optimization~\cite{ 9625086 }.

The Emulation stage offers a realistic preview of experiment performance. Experimenters refine and verify their experiments here, which are then executed by AERPAW operators on the outdoor testbed.\\

Key components of AERPAW's DT include:
\begin{itemize}
    \item \textbf{Virtual USRPs}: are simulated USRP devices interacting through the USRP hardware driver (UHD), enabling radio configuration testing without physical hardware.
    \item \textbf{Channel Emulator}: emulates real-world signal propagation phenomena, providing realistic radio channel conditions for accurate testing.
    \item \textbf{Vehicle Emulation}: integrates with the channel emulator to reflect the dynamic movement and orientation of virtual drones in the emulated environment~\cite{aerpawemulation}.
\end{itemize}

AERPAW's DT is structured into several layers to emulate different aspects of wireless communications:

\begin{itemize}
    \item \textbf{Packet Layer Emulation}: uses AERPAW CHEM-VM integrated with Zero Message Queue for SDR communication over the IP network, simulating packet data traversal and channel matrix computations.
    \item \textbf{IQ Layer Emulation}: employs UHD on a virtual USRP (V-USRP) to simulate in-phase and quadrature signals 
    in a controlled environment.
    \item \textbf{RF Layer Emulation}: utilizes a physical USRP connected through PROPSIM, a radio channel emulator that replicates real-world wireless conditions by simulating various RF channel attributes.
\end{itemize}

Experiments designed within the DT can be seamlessly transferred to AERPAW's outdoor testbed for real-world execution. Once conducted, results are sent back to the users for analysis. While DTs offer significant benefits such as scalability, cost-effectiveness, and flexibility, AERPAW's DT is complemented by hardware-in-the-loop setups (Sandbox) and the outdoor testbed~\cite{10.1145/3396865.3398692}. These setups provide realistic operational environments essential for final validation and comprehensive data collection.

DTs excel in initial development and iterative testing, though they may have limitations due to simplified models of wireless channel propagation and physical antenna patterns. Therefore, physical testbeds remain indispensable for validating results and gathering real-world data.

\section{{\textcolor{black}{Experimental procedures and Results}} 
}
\label{sec:implementation}
\begin{figure}
    \centering
    \includegraphics[width=1\linewidth]{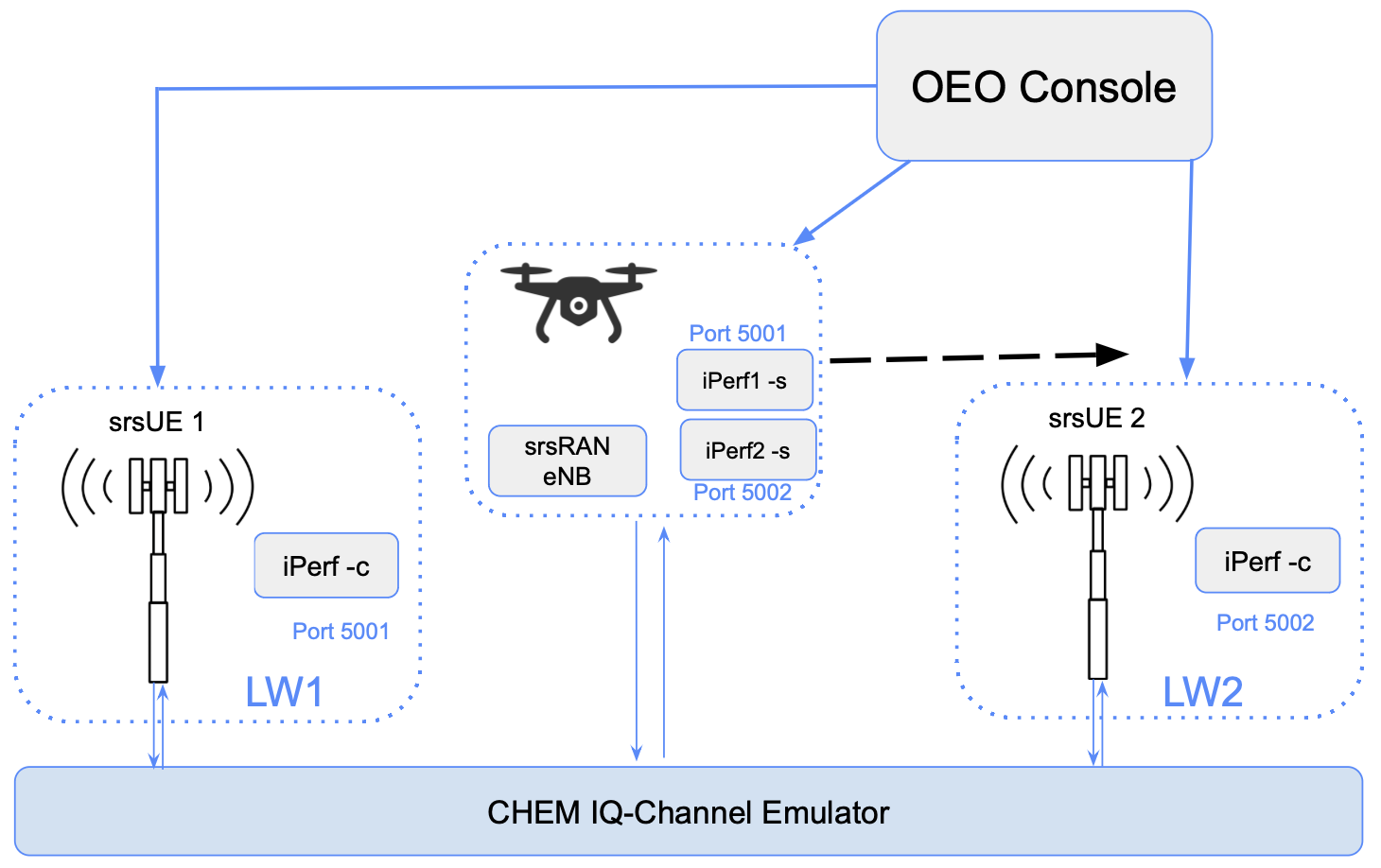}
    \vspace{-18 pt}
    \caption{Aerial base station serving fixed ground UEs.}
    \label{fig:experimentoverview}
    \vspace{-4mm}
\end{figure}

\begin{figure}
    \centering
    \includegraphics[width=1\linewidth]{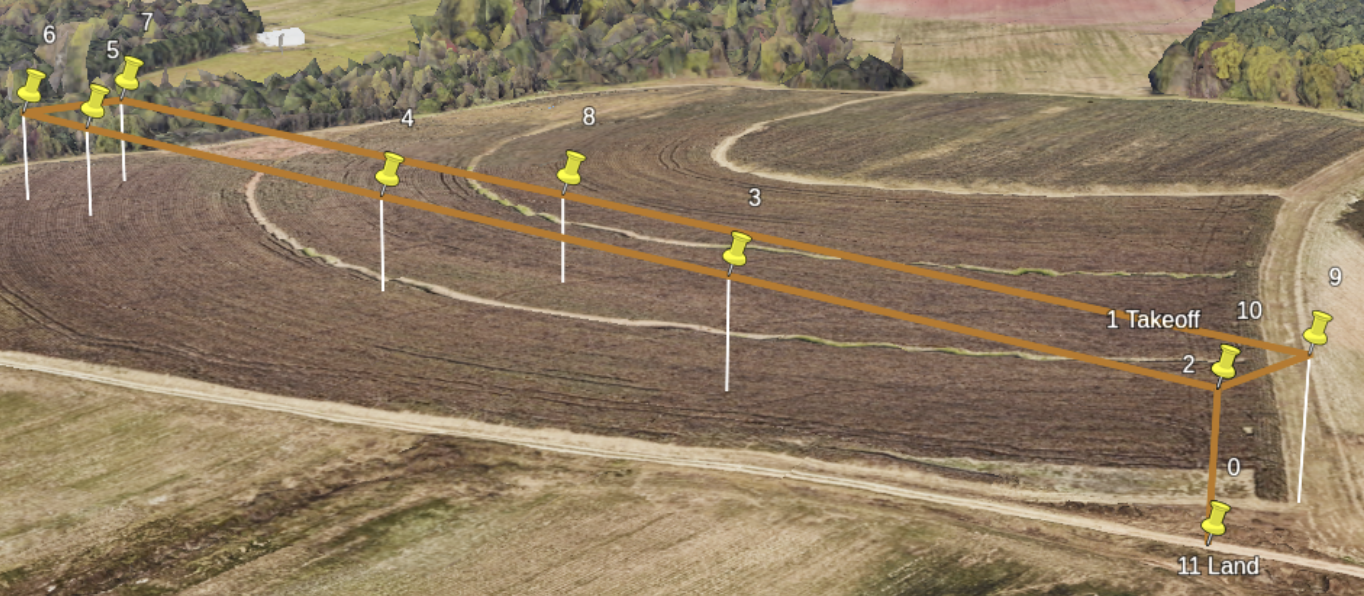}
    \vspace{-15 pt}
    \caption{Custom flight path from LW1 to LW2.}
    \label{fig:flightpath}
\end{figure}

\subsection{Experimental Approach}

We utilize AERPAW's DT to simulate a connection between two V-USRPs, an srsRAN eNodeB (eNB), and two User Equipment (UE), running version 23.4.0 of the srsRAN software. In this scenario visualized in Fig.~\ref{fig:experimentoverview}, an aerial eNB mounted on an UAV will serve two fixed ground nodes while following a predefined flight path, starting from one fixed node and flying towards the other. This experiment provides insights into UAV based communications systems operating under various constraints, such as limited battery life, dynamic changes in signal strength, and connectivity. The setup and execution of the experiment on the AERPAW platform are described in detail in continuation, and are illustrated in Fig.~\ref{fig:experimentation}.

\begin{figure}[t]
    \centering

    \includegraphics[width=0.85\columnwidth]{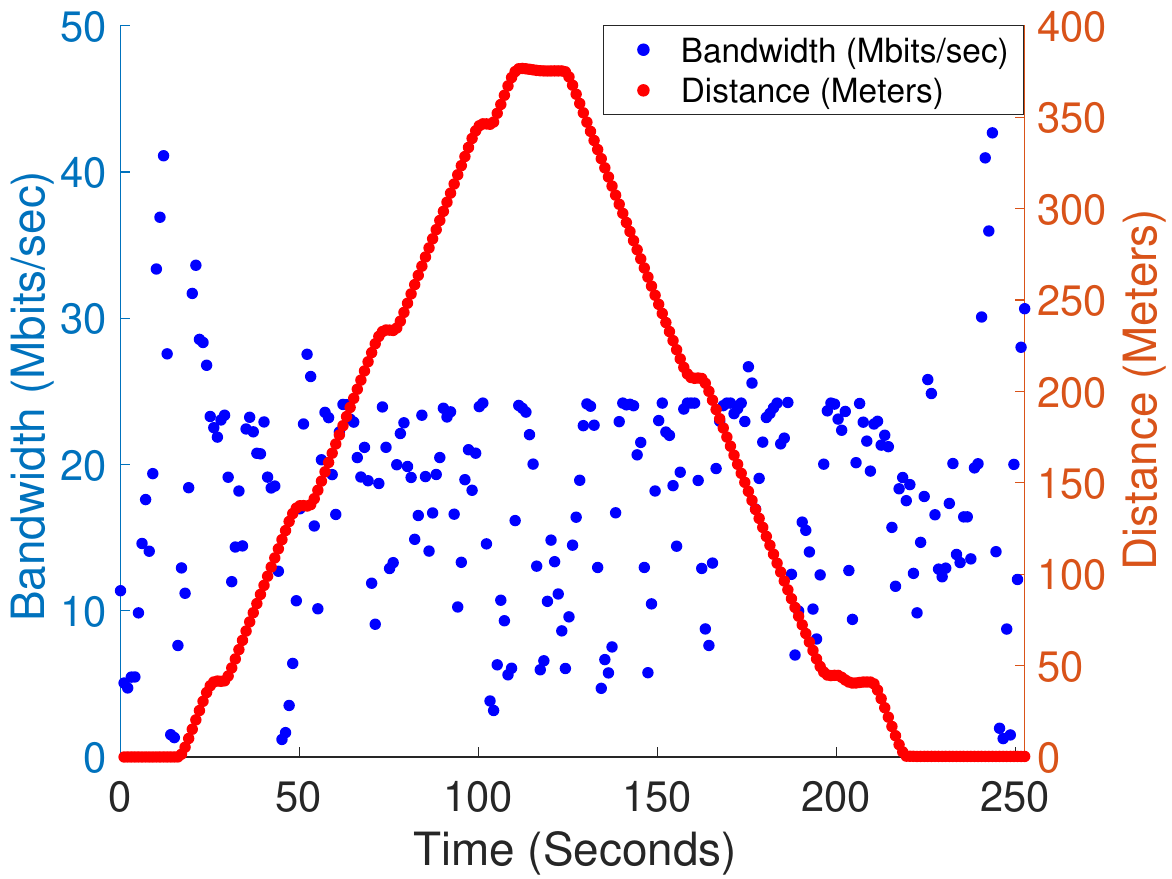}
    \vspace{-5 pt}
    \caption{Achieved throughput in Mbps plotted as blue scattered (left Y-axis) and distance from LW1 in meters plotted as red line (right Y-axis) over time (X-axis). }
    \label{fig:lw1}
\end{figure}

\begin{figure}[t]
    \centering
    \includegraphics[width=0.85\columnwidth]{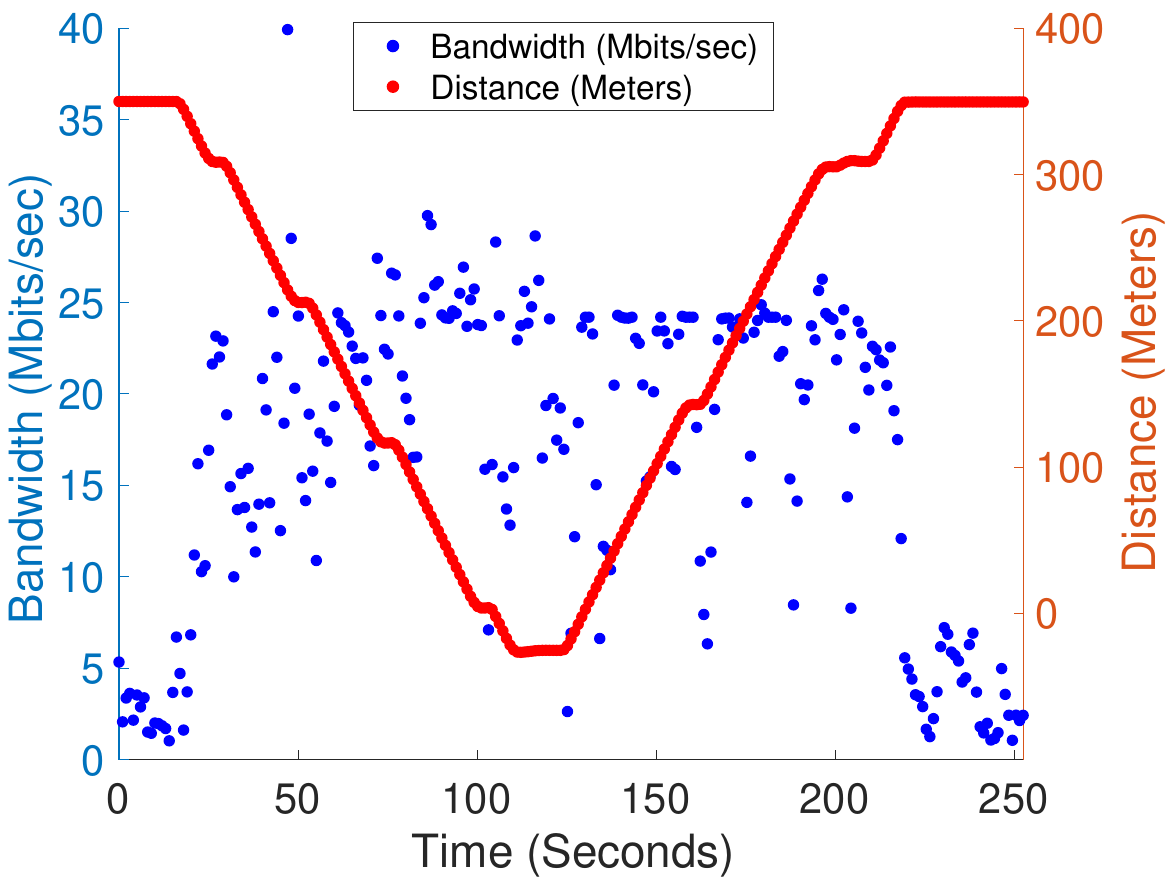}
    \vspace{-5 pt}
    \caption{Achieved throughput in Mbps plotted as blue scattered (left Y-axis) and distance from LW1 in meters plotted as red line (right Y-axis) over time (X-axis). 
    } 
    \label{fig:lw2}
    \vspace{-5mm}
\end{figure}

\begin{enumerate}
    \item \textbf{Configuration and Setup}: Initially, we configure the experiment by allocating resources at Lake Wheeler Fixed Node 1 (LW1), Lake Wheeler Fixed Node 2 (LW2), and the Large AERPAW Portable Node (LPN1) used on the UAV. ``Large'' in this context refers to a slight increase in computational power but also increased weight, reducing UAV flight time in physical experiments. We then download the necessary virtual private network (VPN) configuration file for secure shell (SSH) container access. Once connected, we access the containers via SSH and set up the scripts required for traffic generation, vehicle positioning, and radio configurations.
    \item \textbf{Script and Configuration Setup}: We modify the vehicle script to reference a custom flight plan file generated using QGroundControl as shown in Fig. \ref{fig:flightpath}, which defines the movement patterns of the virtual vehicles in the emulator. LW1 is at point 0 and LW2 is at point 6 in the figure. This flight path has the UAV takeoff from LW1 and fly to LW2 at a fixed velocity. Once it reaches LW2 it then flies back to LW1. 
    \item \textbf{Execution}: The OEO console monitors processes and allows the experimenter to issue commands to the vehicle, facilitating real-time control and adjustments. The experiment is launched using the \textit{startexperiment.sh} script, initiating all processes and beginning data collection. During the experiment, logs are monitored in real-time to ensure proper operation and address any issues. Upon completion, the \textit{stopexperiment.sh} script is used to stop all processes, and the \textit{reset.sh} script is executed to clear residual configurations and prepare for subsequent experiments.
    \item \textbf{Post-Processing and Analysis}: Collected logs are processed using AERPAW’s Python scripts. Specifically, \textit{akmlGen.py} generates KML files for geospatial data visualization, \textit{log2csv.py} converts logs to CSV format, \textit{plotCsv.py} creates plots from CSV data, and \textit{csvMerge.py} consolidates multiple CSV files if necessary. Results are visualized using MATLAB for detailed data analysis, \textit{Matplotlib} for generating plots, and Python’s \textit{simplekml} library for handling Keyhole Markup Language (KML) files, enabling a comprehensive analysis of the experiment's outcomes. Once the experiment completes, 
    we use a combination of our own scripts and AERPAW's post processing scripts to process and visualize the data.
    
\begin{figure}
    \centering
    \includegraphics[width=1\linewidth]{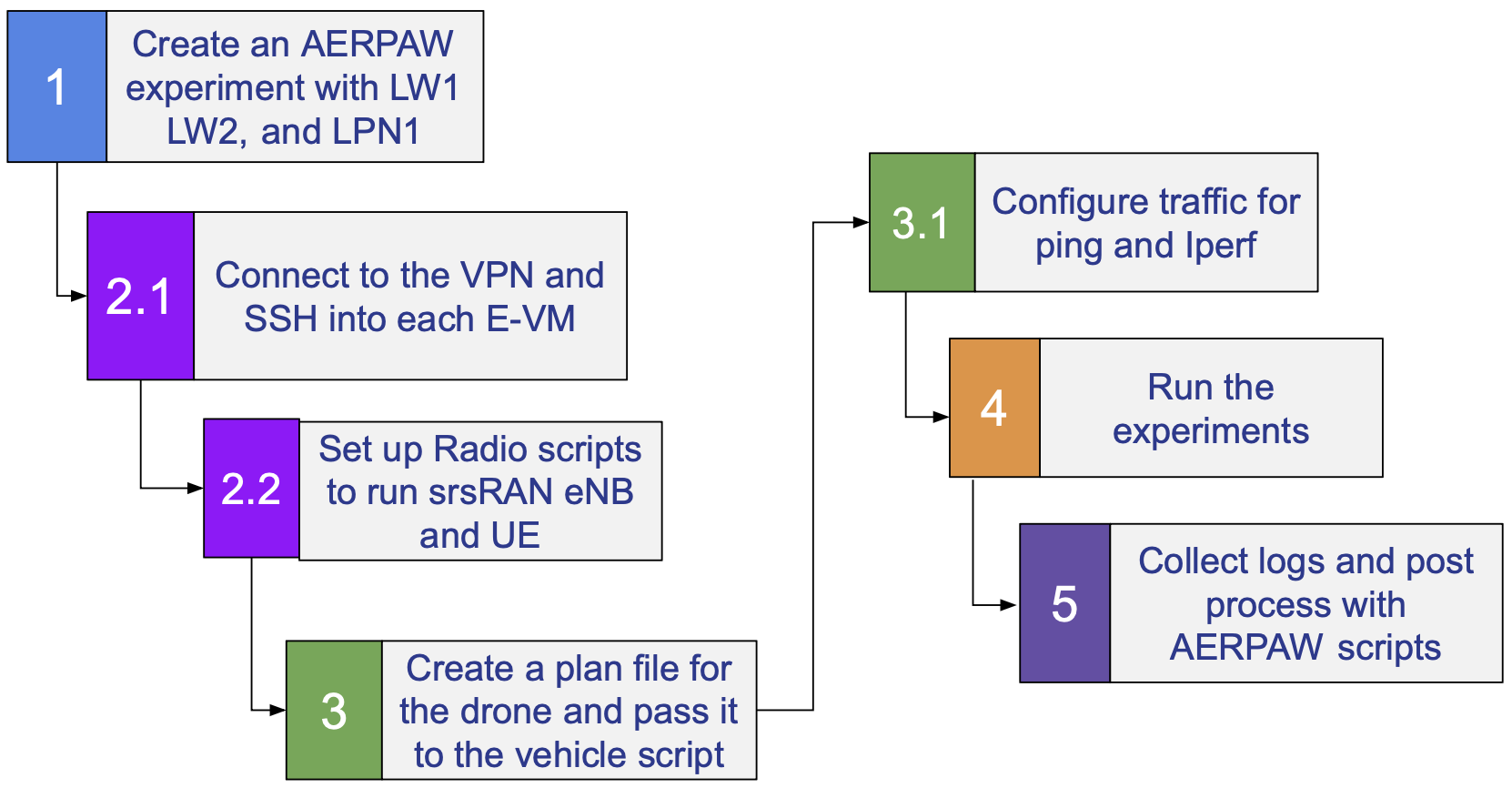}
    \vspace{-15 pt}
    \caption{High level process of experimentation.}
    \label{fig:experimentation}
    \vspace{-5mm}
\end{figure}
\begin{enumerate}
    \item \textbf{Processing Logs into CSV}: When an AERPAW experiment is run, unique timestamped logs are produced for each process. We use AERPAW-provided scripts to parse the logs and collect necessary data. 
    \item \textbf{KML Generation}: To visualize the collected data relative to the emulated position of the UAV (altitude, longitude, and latitude), AERPAW uses Python \textit{simplekml} to create color-coded map data for analysis. Data on drone location and a user-specified metric, e.g., channel bandwidth, is loaded into the KML format. This XML-based file format can be loaded into Google Earth or other software to view data overlaid on a map.
    \item \textbf{Plotting with Matplotlib and Matlab}: We create custom Python scripts to visualize the relationship between distance and link quality. Since KML files are limited in their ability to represent the actual quality of a communications channel, we implement several approaches to plotting. As shown in Figs.~\ref{fig:lw1} and \ref{fig:lw2}, we graph the distance from the UE using a haversine function, then overlay a 
    bandwidth plot alongside it.
\end{enumerate} 

\end{enumerate}

\begin{figure}[t]
    \centering
    \includegraphics[width=0.85\columnwidth]{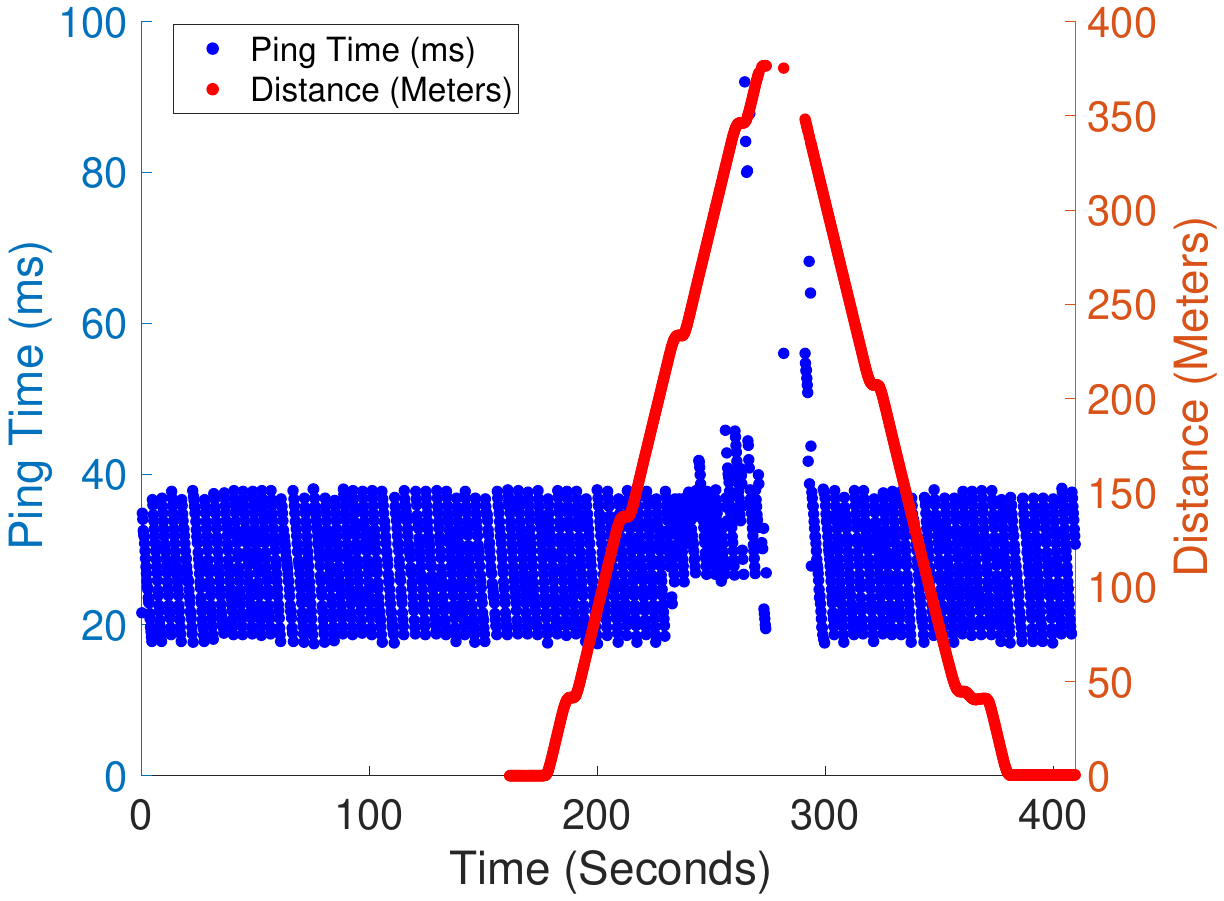}
    \vspace{-5 pt}
    \caption{Measured pingtime plotted as blue scattered in ms (left Y-axis) and distance from LW1 in meters plotted as red line (right Y-axis) over time (X-axis).}
    \label{fig:pingLW1}
\end{figure}

\begin{figure}[t]
    \centering
    \includegraphics[width=0.85\columnwidth]{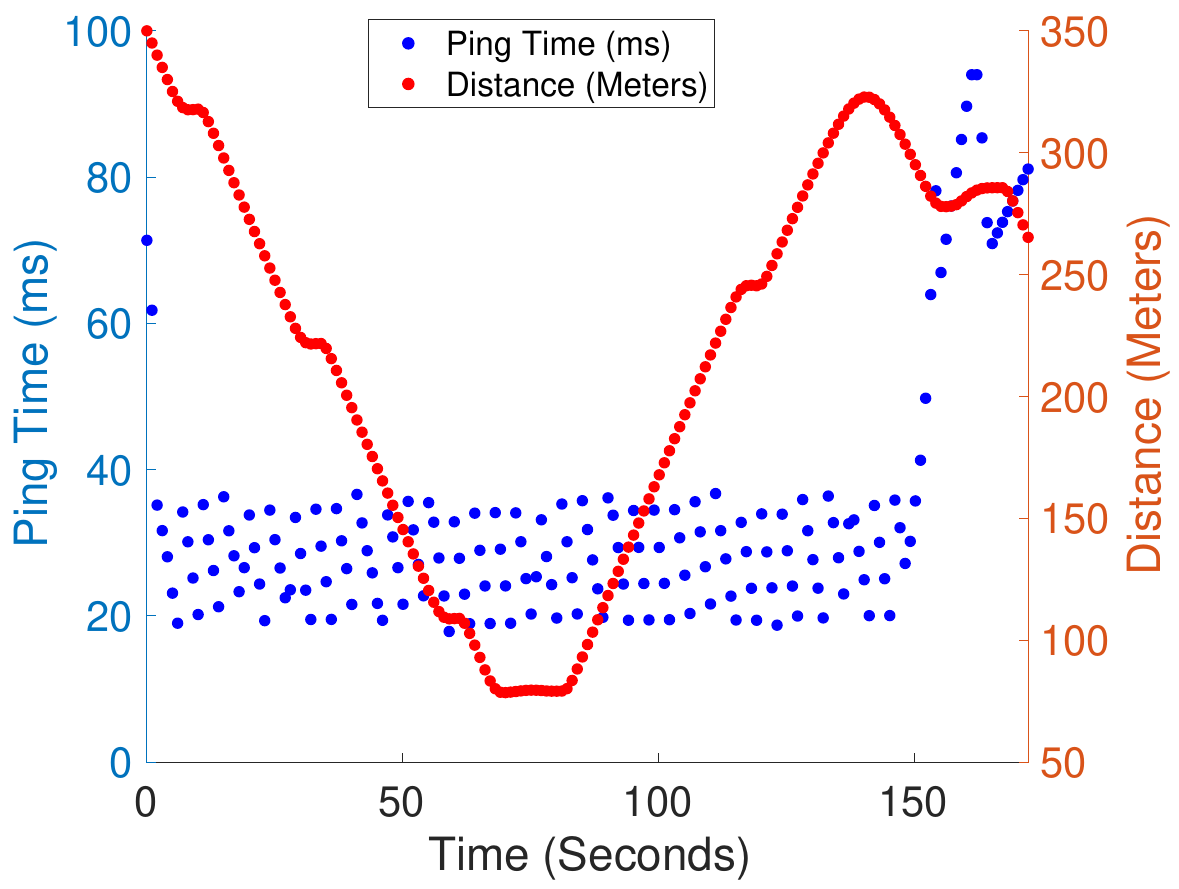}
    \vspace{-5 pt}
    \caption{Measured pingtime plotted as blue scattered in ms (left Y-axis) and distance from LW1 in meters plotted as red line (right Y-axis) over time (X-axis).}
    \label{fig:pinglw2}
\end{figure}

\subsection{Results and Analyses}
The results of the experiments are shown in Figs.~\ref{fig:lw1} and \ref{fig:lw2} 
which shows the throughput achieved over distance. The measurement provides an opportunity to observe how an aerial base station can support multiple ground UEs and how distance affects this connection. Fig. \ref{fig:flightpath} shows the trajectory of the UAV which starts at LW1 and flys to LW2. All results  presented in this paper follow this same flight path. Both UEs at LW1 and LW2 are connected to the base station and measured simultaneously. Using Iperf, we simulate traffic to each user, allowing us to observe how the srsRAN scheduler allocates resources and ensures fair distribution of bandwidth between the two UEs. The default scheduler in srsRAN is a round-robin scheduler. This scheduler distributes resources evenly among all active users without prioritizing any specific user or traffic type, ensuring a fair allocation of available bandwidth. 

In a Time-Division Duplexing system with an evolved NodeB (eNB) mounted on a UAV and using a round-robin scheduler, the throughput for each UE can be analyzed mathematically. In a round-robin scheduling scheme, resources are allocated sequentially to users in a cyclic order. For two users, the scheduler allocates resources to each user in turn, ensuring that each user receives an equal share of the available bandwidth over time. It is a 20 MHz channel with 100 Resource Blocks (RBs) and either 64-QAM or 16-QAM modulation schemes, the throughput calculation is as follows. For 64-QAM, where each symbol carries 6 bits, the number of bits per RB is $\text{Bits per RB}_{64\text{-QAM}} = 84 \text{ symbols} \times 6 \text{ bits/symbol} = 504 \text{ bits}$. Thus, the total number of bits for 100 RBs is $\text{Total Bits}_{64\text{-QAM}} = 504 \text{ bits/RB} \times 100 \text{ RBs} = 50,400 \text{ bits}$. With round-robin scheduling, each user receives half of the total capacity. The average throughput when 64-QAM is used is given by dividing 54 Mbits among the two users, which results in 25.2 Mbps. Following the same analysis for the 16-QAM with 4 bits per symbol, the $\text{Average Throughput}_{16\text{-QAM}}$ is $16.8 \text{ Mbps}$.



These calculations are helpful to show the difference in achieved average throughput as the channel conditions worsen with distance as shown in Figs. \ref{fig:lw1} and \ref{fig:lw2}. Thus, switching from 64-QAM to 16-QAM reduces the throughput per user from 25.2 Mbps to 16.8 Mbps as shown in Fig. \ref{fig:lw1} when the UAV is furthest from LW1. Inversely in Fig. \ref{fig:lw2} the optimal throughput is achieved once the UAV is closest to LW2. This reduction is due to the lower bit capacity per symbol in 16-QAM compared to 64-QAM, reflecting a trade-off between achieving higher data rates and maintaining robustness against poorer channel conditions.
Iperf is showing the network performance metrics between the base station and the UEs, specifically the 
data transfer rates. Figures~\ref{fig:pingLW1} and \ref{fig:pinglw2} plot the ping results from both LW1 and LW2 to the eNB at the same time the UAV is making its way through the predefined flight path. Using ping to measure latency between an aerial eNB on a UAV and a fixed ground UE helps evaluate the responsiveness of the wireless connection by determining the ping time, which is the round-trip time it takes for data packets to travel between the two nodes. This measurement provides insights into network latency performance, which is critical for maintaining reliable communication in UAV-based systems. We can see that the latency is worst when the UAV is furthest from the eNB with a brief disconnect occurring in Fig. \ref{fig:pingLW1} at the furthest distance from LW1. Further analysis of both pingtime figures reveals an acceptable threshold for distance being around 150 meters before the increase in latency starts to worsen.  




\section{Conclusions}
\label{sec:conclusions}
The results of our research indicate the potential of AERPAW's DT, built along the AERPAW testbed, in advancing UAV communication technologies. By leveraging AERPAW's DT and vehicle emulator, one can emulate a wide range of communications scenarios, 
providing valuable insights into system performance and reliability that are difficult to achieve through theoretical models alone. Our experiments involved configuring and executing scripts on the AERPAW platform, monitoring logs, and using advanced visualization tools 
to interpret the data. This comprehensive approach not only validates the effectiveness of the AERPAW DT but also highlights its utility in bridging the gap between simulation and real-world outdoor experiments. 
Moving forward, continued research should focus on refining DT models, improving emulation accuracy, and exploring more complex scenarios to fully exploit their potential. By integrating DTs with real-world testbeds, the UAV research community can achieve a more profound and practical understanding of UAV systems, driving innovation and facilitating more efficient deployment across various applications.

\section*{Acknowledgement}
This work was supported in part by the NSF PAWR program, under grant number CNS-1939334. 

\balance

\bibliographystyle{IEEEtran}
\bibliography{aly}





\end{document}